\documentstyle[12pt,aaspp4,epsfig]{article}

\begin{document}

\title{%
Optical depth evaluation in pixel microlensing 
}

\author{Paolo Gondolo}
\affil{Max Planck Institut f\"ur Physik, F\"ohringer Ring 6,
  80805 Munich, Germany\\{\rm Email: \tt gondolo@mppmu.mpg.de}}

\begin{abstract}
  We propose an estimator of the microlensing optical depth from pixel lensing
  data that involves only measurable quantities.  In comparison to the only
  previously proposed estimator, it has the advantage of not being limited to
  events with large magnification at maximum, and it applies equally well to
  satellite and ground-based observations.
\end{abstract}


\vfill\eject

\section{INTRODUCTION}

An important physical quantity that we want to determine from gravitational
microlensing data is the microlensing optical depth. This is the number of
gravitational lenses lying within one Einstein radius around the
line of sight to a given source.

When monitoring individual stars as proposed by Paczy\'nski
\markcite{pac86}(1986), the optical depth is estimated from observational data
as a weighted sum of the Einstein times of the individual events (see
eq.~[\ref{usual}] below; Griest \markcite{gri91}1991).  The Einstein times 
are determined by fitting the theoretical microlensing
lightcurve to the measured lightcurve. 
The fit is almost degenerate in the
three fit parameters Einstein time, maximum magnification, and star flux in
absence of lensing (Gould \markcite{gou96}1996; Wo\'zniak \& Paczy\'nski
\markcite{woz97}1997). In particular, the fitted value of the Einstein time depends
sensitively on the value of the star flux in absence of lensing, especially
at large maximum magnifications.

In pixel microlensing, which is gravitational microlensing of unresolved stars,
the star flux in absence of lensing is not generally
accessible, since the combined light of all the stars contributing to a pixel is monitored
simultaneously. 
Thus, because of the above--mentioned degeneracy, the Einstein time
cannot be reliably determined from a fit to the lightcurve.  In this case, the
optical depth cannot be estimated by means of the usual expression, and this seems
to hinder the possibility of obtaining it from pixel lensing
data.

A partial solution to this problem has been proposed by Gould
\markcite{gou95}(1995 \& \markcite{gou96}1996) under conditions that may be possible in
future satellite observations but are definitely not typical of present
microlensing searches towards the Andromeda galaxy M31 and the Large Magellanic
Cloud (Crotts \markcite{cro92}1992, Baillon et al.  \markcite{bai93}1993,
Crotts and Tomaney \markcite{cro96}1996, Tomaney and Crotts
\markcite{tom96}1996, Ansari et al.  \markcite{ans97}1997 for M31; Melchior et
al. \markcite{mel97}1997 \& \markcite{mel98}1998 for the LMC). He assumed
regular observations with constant exposure times, a constant level of noise
dominated by photon counting, and large magnifications at maximum so that the
microlensing lightcurve assumes a completely degenerate form
(eq.~[\ref{degenerate}] below). The present pixel--lensing searches instead
have irregular time gaps between observations, show a variable level of noise
depending on factors other than photon counting, and expect relatively low
magnifications for which the microlensing curve is not well approximated by the
completely degenerate form.

In this letter, we present an estimator of the optical depth in pixel
microlensing that applies to all magnifications, involves measurable quantities
only, and is of direct applicability to the data.  It is based on the
substitution of the Einstein time with the product of the flux increase at
maximum and of the full event duration at half-maximum,
defined as the total time during which the flux increase exceeds half of its
maximum value. 

\section{USUAL OPTICAL DEPTH ESTIMATOR}

The flux increase in a microlensing event is given by (for pointlike lens and
pointlike source; Einstein \markcite{ein36}1936)
\begin{equation}
  \Delta F(t) = F \, f(u^2(t)) ,
\end{equation}
with
\begin{equation}
\label{eq:f}
  f(x) = \frac{ 2 + x }{\sqrt{ x ( 4 + x )}} - 1 .
\end{equation}
Here $F$ is the source flux in absence of lensing, $u=\theta/\theta_e$, 
 $\theta$ is the angular separation of lens and
source, and $\theta_e$ is the angular Einstein radius
\begin{equation}
  \theta_e = \left[ \frac{4 G m}{c^2} \, \frac{ s-l}{s l} \right]^{1/2},
\end{equation}
where $m$ is the lens mass, and $s$ and $l$ are the distances of the source
and the lens from the observer.

The relative motion
of lens and source during the event is usually well approximated by a uniform
motion, for which
\begin{equation}
  \label{microcurve2}
  u^2(t) = \beta^2 + \left( \frac{t-t_{\rm max}} {t_e} \right)^2 .
\end{equation}
$t_{\rm max}$ is the time of maximum magnification, $\beta$ is the impact
parameter in units of the Einstein radius, and $t_e$ is the Einstein time. This
we define as the time the lens would take to cross the {\it radius} of the
Einstein ring, $t_e = \theta_e/\mu$, where $\mu$ is the relative proper motion
of lens and source.

In the limit of large magnifications at maximum, the
microlensing curve assumes the completely degenerate form
\begin{equation}
  \label{degenerate}
  \Delta F(t) \simeq {F \over u(t) } = {F \over \beta} \left[ 1 +
  \left(t-t_{\rm max} \over \beta t_e \right)^2 \right] ^ {-1/2} \qquad\qquad
    (F, \beta \to 0; \, F/\beta = {\rm const}) .
\end{equation}
The three parameters $F$, $\beta$ and $t_e$ have reduced to two: $F/\beta$ and
$\beta t_e$. A fit to the data can determine neither the star flux $F$ nor the
Einstein time $t_e$ separately from the unknown impact parameter $\beta$ (see
e.g.\ Wo\'zniak \& Paczy\'nski \markcite{woz97}1997). 

The microlensing optical depth is the number of gravitational
lenses lying within one Einstein radius $\theta_e$ around the line of sight to
a given source,
\begin{equation}
\label{optdepth}
  \tau = \int_0^s \pi \theta_e^2 l^2 \, n(l) \, dl,
\end{equation}
where the integral is along the line of sight to the source at distance $s$
from the observer, and $n(l)$ is the lens number density at distance $l$ along
the same line of sight.

When monitoring $N_s$ sources for a time $T$, the optical depth is usually
estimated as 
\begin{equation}
  \label{usual}
  \overline{\tau} = {\pi \over 2 N_s T} \sum_{\rm events} { t_{e} \over
    \epsilon(t_{e}) } .
\end{equation}
Here $\overline{\tau}$ is the optical depth averaged over the monitored
sources, the sum is over the observed events, and $\epsilon(t_e)$ is the
detection efficiency (normalized to impact parameters $<\theta_e$) as a
function of Einstein time $t_e$. The Einstein time of each event is determined
experimentally by a fit of the theoretical microlensing lightcurve to the
measured lightcurve.

This usual estimator is based on the relation
\begin{equation}
\label{basic_usual}
\int_{\beta<1} t_e dN_s d\Gamma = \frac{\pi}{2} N_s \overline{\tau} .
\end{equation}
In the right hand side, $\overline{\tau} = N_s^{-1} \int \tau dN_s$; in the
left-hand side, $ d N_s $ is the differential number of monitored stars, $ d
\Gamma $ is the event rate per source, $ d \Gamma = (2 \tau/\pi t_e) \, d\beta
$, and the integration is limited to impact parameters $\beta < 1$.
Eq.~(\ref{usual}) follows from solving eq.~(\ref{basic_usual}) for
$\overline{\tau}$ and rescaling the observed number of events $ d N^{\rm obs}$
to the expected number of events $ d N = T dN_s d\Gamma$ by means of the
detection efficiency $\epsilon(t_e)$ normalized to $\beta < 1$. Explicitly
\begin{equation}
\sum_{\rm events} \frac{t_e}{\epsilon(t_e)} = \int_{\beta<1}
\frac{t_e}{\epsilon(t_e)} d N^{\rm obs} = \int_{\beta<1} t_e dN = \frac{\pi}{2}
T N_s \overline{\tau} .
\end{equation}

\section{NEW OPTICAL DEPTH ESTIMATOR}

We now prove a relation similar to eq.~(\ref{basic_usual}) but not involving
the Einstein time $t_e$, which is very poorly measurable in pixel lensing.  Let
$t_{\rm fwhm}$ be the event duration at half-maximum and $\Delta_{\rm max}$ be
the flux increase at maximum. Both are easily determined from a fit to the
lightcurve. Then
\begin{equation}
\label{basic_ours}
\int t_{\rm fwhm} \Delta_{\rm max} dN_s d\Gamma = \frac{\pi}{2} I F_s
\overline{\tau}_F .
\end{equation}
Here $ F_s $ is the total flux of the monitored sources, $ I $ is 2.1465, and
$\overline{\tau}_F$ is the optical depth averaged over the monitored sources,
each source weighted by its flux, $\overline{\tau}_F = F_s^{-1} \int F \tau
dN_s$.  This flux-weighted optical depth $\overline{\tau}_F$ equals the
averaged optical depth $\overline{\tau}$ whenever the source luminosity
function is, or can be considered as, independent of source position, because
in this case the averages over source positions and fluxes are independent.
This is the case for a single source population.\footnote{The flux-weighted
  optical depth may actually find an interesting application for individually
  monitored sources, for example in the microlensing data towards the galactic
  bulge: comparing it to the unweighted optical depth may give information on a
  spatial segregation of the star populations.}

The proof of eq.~(\ref{basic_ours}) is as follows.  We express the half-maximum
duration $t_{\rm fwhm}$ as a function of $t_e$ and $\beta$ by finding the value
of the impact parameter $\beta_w$ at which the flux increase equals half of its
maximum value. We also express the maximum flux increase $ \Delta_{\rm max} $
in terms of the source flux $F$ in absence of lensing and of the impact
parameter $ \beta $. We find $ t_{\rm fwhm} = t_e w(\beta) $ and $ \Delta_{\rm
  max} = F \delta(\beta)$, with $ w(\beta) = 2 \, \left( \beta_w^2 - \beta^2
\right)^{1/2} $, $ \beta_w^2 = 2 f(\delta(\beta)) $, $ \delta(\beta) =
f(\beta^2) $, and $f(x)$ given in eq.~(\ref{eq:f}).  The integral over $\beta$
in eq.~(\ref{basic_ours}) is then
\begin{equation}
I \equiv \int_0^\infty w(\beta) \delta(\beta) d\beta = 2.1465 .
\end{equation}
The remaining integral over the sources gives the total source flux $ F_s =
\int F dN_s $ times the flux-weighted optical depth
$\overline{\tau}_F$. Putting the factors together gives eq.~(\ref{basic_ours}).

If all events could be detected, then eq.~(\ref{basic_ours}) would immediately
give an estimator for the optical depth, namely
\begin{equation}
\label{ours_ideal}
\overline{\tau}_F = \frac{\pi} {2 I T F_s} \, \sum_{\rm all~events} t_{\rm
  fwhm} \Delta_{\rm max} \qquad\qquad \hbox{(no noise, 100\% efficiency).}
\end{equation}
Notice that apart from the numerical factor $\pi/2I \simeq 0.73$,
$\overline{\tau}_F$ equals the sum over events of the products of the
fractional flux increase at maximum $\Delta_{\rm max}/F_s$
and of the fraction of observing time spent above half-maximum $t_{\rm
fwhm}/T$. The estimator in eq.~(\ref{ours_ideal}) does not depend on the source
luminosity funciton. 

In practice, there is always a minimum detectable $\Delta_{\rm max}$, if only
because of Poissonian fluctuations in photon counting. Moreover, the
irregularity and the discreteness of the time coverage limit the range of
event time scales a microlensing search can access.  

We can restrict the sum in eq.~(\ref{ours_ideal}) -- and the integral in
eq.~(\ref{basic_ours}) -- to events with $\Delta_{\rm max} > \Delta $, a
convenient detection threshold.  We obtain
\begin{equation}
\label{ours_thr}
\overline{\tau} = \frac{\pi} {2 I T F_{\rm eff}(\Delta)} \, \sum_{\rm
  events \atop \Delta_{\rm max}>\Delta} t_{\rm fwhm} \Delta_{\rm max},
\end{equation}
where we have defined an effective flux
\begin{equation}
F_{\rm eff}(\Delta) = I^{-1} \, \int dF \phi(F) F \int_0^{\beta_{\Delta/F}}
d\beta w(\beta) \delta(\beta) ,
\end{equation}
$\phi(F)$ being the source luminosity function and $\beta_{\Delta/F} = 2
f(2\Delta/F) $.  This is the basic form of the new optical depth estimator that
we advocate for pixel lensing.  Eq.~(\ref{ours_thr}) is written for a single
source population; for several populations, similar equations apply to each
population separately.

In the next section we discuss the dependence of the effective flux $F_{\rm
  eff}(\Delta)$ and of the new optical depth estimator on the source luminosity
function.

\section{EFFECTIVE FLUX AND SENSITIVITY}

The effective flux $F_{\rm eff}(\Delta)$ is smaller than the total flux $F_s$
and is equal to $F_s$ when $\Delta=0$. It depends in principle on the details
of the source luminosity function, but in practice the ratio $F_{\rm
  eff}(\Delta)/F_s$ is an almost universal function of the fluctuation
magnitude $\overline{m}$ (defined in Tonry \& Schneider \markcite{ton88}1988).
This is exemplified in fig.~1 for four luminosity functions: (1) the local
neighborhood in the $V$-band (Jahrei\ss\ \& Wielen \markcite{jah98}1998); (2)
47 Tucanae in the $V$-band (Hesser et al. \markcite{hes87}1987); (3) the M31
bulge in the $K$-band (Rich, Mould, and Graham \markcite{ric93}1993); (4) the
galactic bulge in the $I$-band (Terndrup, Frogel and Whitford
\markcite{ter90}1990; Holtzman et al.  \markcite{hol98}1998).  The upper panel
shows $\mu_{\rm eff} - \mu $ as a function of $m_{\rm thr} - \overline{m} $,
where $\mu_{\rm eff}$ is the surface magnitude corresponding to the effective
flux, $\mu$ is the actual surface magnitude, and $m_{\rm thr}$ is the magnitude
corresponding to the threshold flux increase $\Delta$. The lower panel shows
the fractional differences in effective flux with respect to the mean over the
luminosity functions considered, which is also the relative error in the
estimated optical depth.  At very low thresholds ($ m_{\rm thr} \gg
\overline{m} $), the effective surface magnitude tends to the actual surface
magnitude, and this for all luminosity functions. At very high thresholds
($m_{\rm thr} \ll m_{\rm tip}$, the magnitude of the brightest star), the
difference between effective and actual surface magnitudes is simply $ \mu_{\rm
  eff} - \mu = \overline{m} - m_{\rm thr} -0.52$, and this again for all
luminosity functions. The largest deviations from a universal behavior occur at
detection thresholds close to the fluctuation magnitude, and for the luminosity
functions considered here they are never larger than 30\%.  If the statistical
errors in the determination of the optical depth will be smaller than these
systematic errors (we recall that the present error on the optical depth
towards the LMC is $ \approx 50\%$ [Alcock et al.\ \markcite{alc97}1997]), the
systematic errors can of course be reduced by measuring the source luminosity
function in deep exposures, even independently of the microlensing search
itself.

A given microlensing search is only sensitive to a range of Einstein times and
radii, and can only measure the contributions to the optical depth coming from
this range. Indeed, with a finite time coverage, there is a minimum event
duration set by the finite exposure time and the interval between successive
images, and a maximum event duration set by the total time span of the
observations. Moreover, for sources with finite angular extensions, there is a
reduction in flux increase with respect to the pointlike case when the lens
transits the source disk. As a consequence, the fraction of observable events
is different according to the Einstein time and radius of each event.

We write the optical depth observable in a given microlensing search as
\begin{equation}
\tau_{\rm obs} = \int \pi \theta_e^2 l^2 n(v_T,l) s(t_e,\theta_e) dv_T dl
\end{equation}
(compare with eq.~[\ref{optdepth}], where $n(l) = \int n(v_T,l) dv_T$). Here
$v_T = l \mu$ is the relative transverse velocity of lens and source. 

The function $s(t_e,\theta_e)$ represents the sensitivity of a given experiment
to the different contributions to the optical depth, and depends on the
observing conditions and data analysis and on the choice of the optical depth
estimator. We call $s(t_e,\theta_e)$ the optical depth sensitivity. For the
usual estimator, $s(t_e,\theta_e)$ equals the detection efficiency as a
function of $t_e$ and $\theta_e$. For the new estimator in
eq.~(\ref{ours_thr}), $s(t_e,\theta_e)$ is
\begin{equation}
\label{sensitivity}
s(t_e,\theta_e) = \frac{1}{I T F_{\rm eff}(\Delta)} \, \int \, F \phi(F) \,
  w(\beta) \delta(\beta) \, \epsilon_{\rm obs}(F,\beta,t_e,\theta_e,t_{\rm
  max}) \, dF d\beta dt_{\rm max} ,
\end{equation}
where $ \epsilon_{\rm obs}(F,\beta,t_e,\theta_e,t_{\rm max}) $ is the event
detection efficiency (counting efficiency).

In an actual survey, the optical depth sensitivity, like the efficiency, has to
be evaluated in a Monte-Carlo simulation. Events are generated, and two
histograms in the binned $(t_e,\theta_e)$ plane are constructed: one counting
the events detected by the selection algorithm in the given observing
conditions, the other including all events above the reference threshold
$\Delta$. Weighting each event by $ F \phi(F) w(\beta) \delta(\beta) $ and
summing over events, the two histograms evaluate the numerator and denominator
of eq.~(\ref{sensitivity}).  The sensitivity is then their bin-by-bin ratio.

To illustrate the dependence of the optical depth sensitivity on the luminosity
function, we consider a simple expression for the detection efficiency. We
count all events for which the flux increase stays above the threshold $ \Delta
$ for a duration longer than $t_1$ but shorter than $t_2$. In addition, in this
example we neglect photometric errors and finite-source effects (so to have a
dependency on $t_e$ only). This translates into $\epsilon_{\rm obs} = 1 $ for $
t_1 < 2 t_e \sqrt{ \beta_{\Delta/F}^2 - \beta^2 } < t_2$ and $\epsilon_{\rm
  obs} =0 $ otherwise. We calculate the optical depth sensitivity for the
luminosity functions mentioned previously, and plot the results in fig.~2. Each
panel corresponds to a value of the threshold magnitude $m_{\rm thr}$, and
$t_2/t_1$ is fixed at 100. Although the details differ from luminosity function
to luminosity function, the sensitivity curves are similar, especially
considering that with a more realistic expression for the detection
efficiency these curves would be smoothed out and that in a comparison with a
model they would be folded with the model $t_e$ distribution and then
integrated.  Again, a measurement of the luminosity function would reduce even
these small systematic uncertainties.

We finally comment on finite-source effects. With sources of finite angular
diameter, the flux increase in a microlensing event does not reach the value of
the pointlike case whenever the lens is over the disk of the source. The exact
light curve depends on the luminosity distribution on the disk of the star, and
in general it is a complicated function. A simple account of finite-size
effects consists in stopping the flux increase when $ u(t) < \theta_s / 2
\theta_e $, where $\theta_s$ is the angular radius of the source. Given the
minimum detectable flux increase $\Delta$, events with $F
\delta(\theta_s/\theta_e) < \Delta $ but $F \delta(\beta) > \Delta $ are
undetected. The efficiency becomes a function of the Einstein radius
$\theta_e$, as does the sensitivity. The optical depth sensitivity as a
function of $t_e$ {\it and} $ \theta_e$ can be calculated from
eq.~(\ref{sensitivity}) once a relation between stellar radius and luminosity
is specified. We have checked that also in this case the behavior of the
sensitivity is almost universal.

\section{CONCLUSIONS}

We have proposed an estimator of the optical depth in pixel microlensing
experiments that involves only measurable quantities. Our new estimator is not
limited to events with large magnifications at maximum and applies equally well
to satellite and ground-based observations.

Our estimator is based on the idea of characterizing the event not by the
Einstein time and the maximum magnification but by the full event duration at
half-maximum and the flux increase at maximum. The latter are easily determined
from a fit to the event lightcurve.

The new estimator of the microlensing optical depth is an efficiency-weighted
sum over events of the products of the fractional flux increase at maximum and
of the fraction of observing time spent above half-maximum.

In the ideal case of no noise and complete time coverage, our proposed optical
depth estimator does not depend on the source luminosity function. In a real
situation, an experiment is only sensitive to events above a threshold and to a
range of Einstein times and radii. We have shown that the sensitivity and the
effective flux depend on the luminosity function in an almost universal way
through the surface brightness fluctuation magnitude. Residual uncertainties
can be removed, if needed, by measuring the luminosity function in deep
exposures.

Our new estimator applies not only to pixel microlensing data but also to
blended and individually monitored sources.  In the latter case, it could be
interestingly compared with the standard estimator based on Einstein times. If
agreement is found, the optical depth estimator we propose may indeed open the
way to the determination of physical parameters from pixel microlensing data.

\acknowledgements

The author acknowledges useful comments and discussions with several
participants to the 4th Microlensing Workshop in Paris, January 1998, where
part of this work was first presented.

\newpage

\figcaption[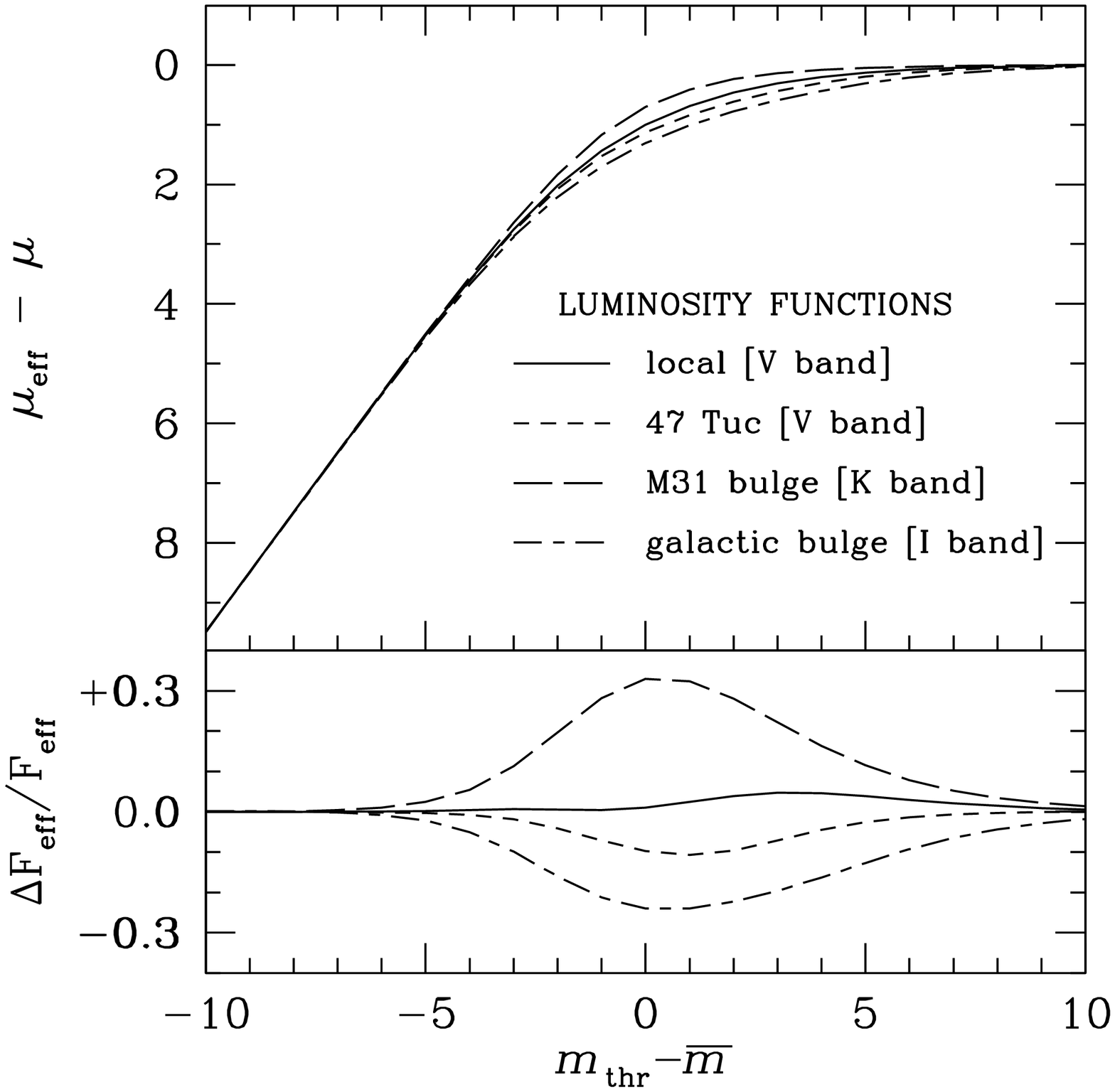]{Dependence of the effective surface magnitude $\mu_{\rm
    eff}$ on the threshold magnitude $m_{\rm thr}$ for various luminosity
  functions. $\mu$ is the actual surface magnitude and $\overline{m}$ is the
  fluctuation magnitude. The lower panel shows the relative systematic error in
  the effective flux with respect to the mean of the luminosity functions
  considered. \label{fig1}}

\figcaption[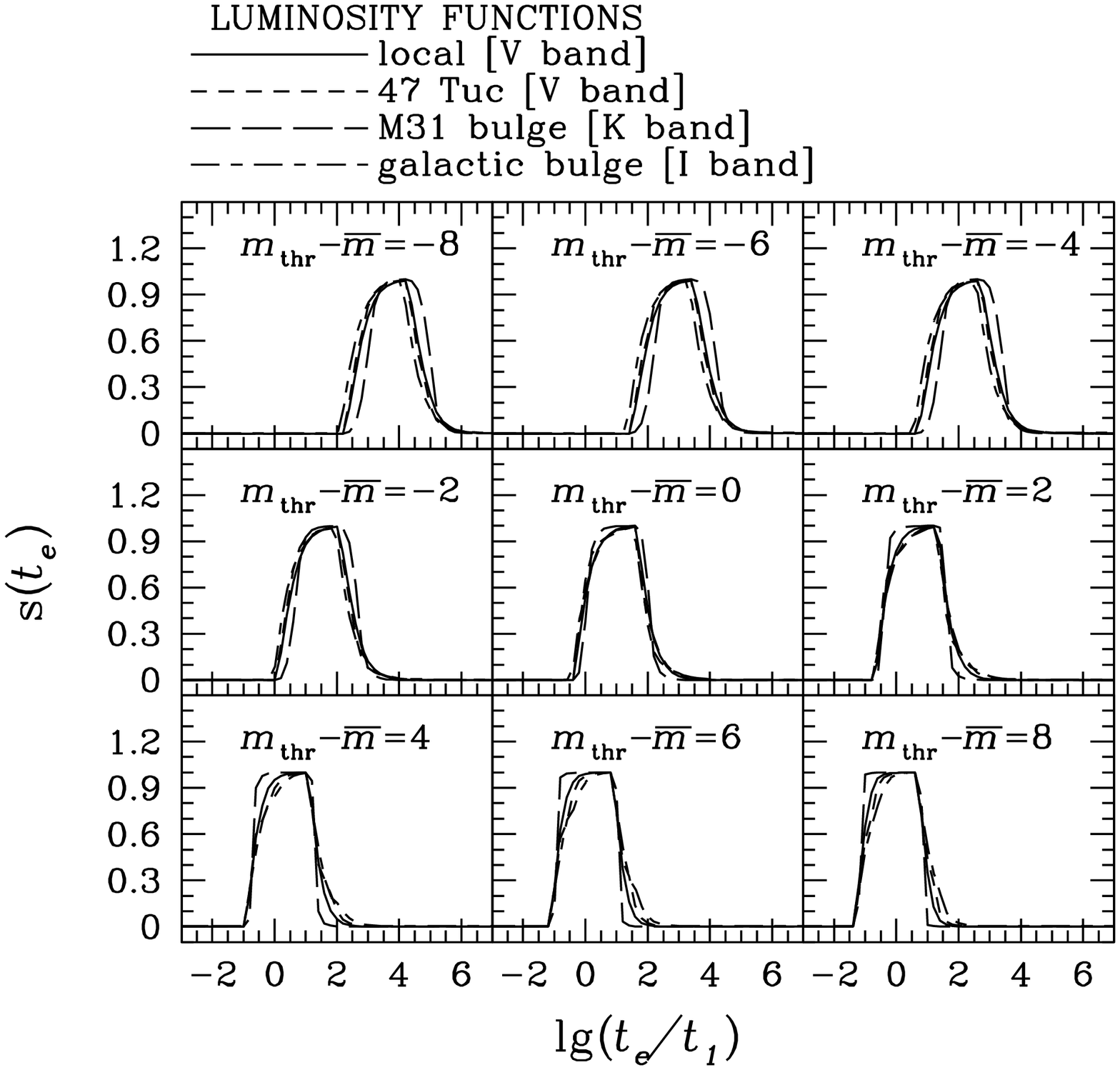]{Optical depth sensitivity as a function of Einstein time
  $t_e$ for a simple detection efficiency and various luminosity functions.
  Each panel corresponds to a different threshold magnitude $m_{\rm thr}$.
  \label{fig2}}

\newpage

\begin{figure}[h]
\centerline{\epsfig{file=odlf1.eps,width=\textwidth}}
\end{figure}

\newpage

\begin{figure}[h]
\centerline{\epsfig{file=odlf2.eps,width=\textwidth}}
\end{figure}

\end{document}